\documentclass[aps, prl, reprint, superscriptaddress, floatfix]{revtex4-2}
\usepackage{graphicx}
\usepackage{dcolumn}
\usepackage{bm}
\usepackage{physics}
\usepackage{amsfonts}
\usepackage{mathrsfs}
\usepackage{subfigure}
\usepackage{color}
\usepackage[svgnames]{xcolor}
\usepackage[colorlinks=true,linkcolor=DarkBlue,anchorcolor=red,citecolor=DarkBlue, urlcolor=DarkBlue]{hyperref}
\DeclareMathAlphabet{\pazocal}{OMS}{zplm}{m}{n}
\begin{document}

\preprint{APS/123-QED}

\title{Electronic glasses from a broken gauge symmetry in disorder-free systems}

\author{Lingyu Yang}
\author{Gia-Wei Chern}
\affiliation{
 Department of Physics, University of Virginia, Charlottesville, Virginia, 22904, USA}

\date{\today}

\begin{abstract}
Glass phases can be stabilized by quenched disorders, as in most spin-glass materials, or self-generated through kinetic freezing in disorder-free systems. A canonical example of the latter is structural glasses, which have been extensively studied for many decades. Yet, how the rugged energy landscape of a glass phase is spontaneously generated in disorder-free systems remains one of the key questions in glass physics. Here we present a general electronic mechanism for the emergence of glassy phase using the example of itinerant electrons coupled to XY spins on a lattice. This model can also be be viewed as the mean-field theory of a superconducting system with attractive density-density interactions. Intriguingly, the electron gauge symmetry in the strong pairing limit gives rise to a macroscopic degeneracy of XY spins. In the presence of electron hopping that breaks the gauge symmetry, the lifting of the extensive degeneracy leads to a glass phase with disordered pairings. Our findings highlight a novel scenario in which a glassy state originates from the breaking of quantum gauge symmetry without quenched disorders. 
\end{abstract}

\maketitle

Glasses are ubiquitous in Nature and are also one of the oldest materials utilized by mankind. Yet despite decades of intensive research, understanding the nature of glass phases remains one of the central subjects in physics. The formulation of glass phenomenology in terms of familiar notions in condensed matter physics often leads to a deeper understanding of glass physics and unexpected connections with other symmetry-breaking phenomena. In particular, several crucial theoretical techniques and ideas, such as glass order parameters and replica symmetry breaking, were originally developed in the context of spin glasses~\cite{binder86,edwards75,sherrington75,parisi86,parisi83,parisi79}.  Spin glasses are magnetic systems where frozen-in disorders give rise to random and frustrated interactions between magnetic moments~\cite{Binder_2003}. The absence of long-range magnetic order in spin glasses is often attributed to the built-in quenched disorder~\cite{sherrington75, Gabay_1981}.

In contrast, glass phases emerge spontaneously in supercooled liquids even without the assistance of built-in disorder~\cite{dyre06,cavagna09,berthier11}. The rigidity in such structural glasses results from the fact that atoms are caged by their neighbors, thus prohibiting their diffusive motions. Glass transitions have even been found in a liquid of hard-sphere particles~\cite{Barrat_1989, Parisi_2005, Pusey_2009, Berthier_2009, Mandal_2014, Xia_2015}, which is perhaps the simplest glass-forming systems. In particular, a unifying picture of glass phases has emerged from the mean-field theory of this canonical model~\cite{parisi10,charbonneau14,maimbourg16,charbonneau17}. It is found that there are actually two distinct glass phases within the amorphous states: a stable glass phase with a simpler energy landscape~\cite{sherrington75, Kirkpatrick_1978}, and a so-called marginal glass characterized by a fractal energy landscape~\cite{Kurchan_2013, Rainone_2015, charbonneau14, Charbonneau_2014}. The two distinct glass phases are seperated by a transition first pointed out by Gardner almost 30 years ago in the study of spin glasses with $p$-spin interactions~\cite{gardner85}. Moreover, the fractal glass phase smoothly merges with the jamming transition in the isostatic limit~\cite{charbonneau17}.  

It is worth noting that, while the above picture is obtained in the limit of infinite dimensions, numerical simulations have shown that several important features persist even in finite dimensions and in other more realistic liquid models~\cite{charbonneau17}. Against the backdrop of these exciting developments, it remains to be seen whether similar scenarios and energy-landscape transitions also occur in other disorder-free glass systems~\cite{Marinari_1994, Bouchaud1994, Cugliandolo1995, Bouchaud1996, Schmalian2000, Westfahl2001, Biroli2001, Tarjus2005}, especially spin-glass type models where several powerful theoretical methods can be applied. Early seminal works have suggested that self-generated randomness are indeed realized in some spin models without quenched disorder~\cite{bouchaud94,Marinari94a,Marinari94b,Lipowski00,Franz01}. A recent work~\cite{yoshino18} on a disorder-free $p$-spin interaction model has shed some light on the nature of self-generated randomness by exploring the mean-field solutions which become exact in the limit of $p\to \infty$ and infinite spin components. Yet, most of these disorder-free spin glass models are rather contrived often with complicated spin interactions.

In this paper, we demonstrate the emergence of a glassy phase in a relatively simple disorder-free XY spin model with effective interactions mediated by itinerant fermions on a lattice. A rugged energy landscape is shown to spontaneously emerge from the breaking of a macroscopic degeneracy that is related to a gauge symmetry. Our work presents a general electronic mechanism for the emergence of glassy states in a disorder-free lattice system. Moreover, the XY spins which reside on the nearest-neighbor bonds of the lattice can be interpreted as superconducting pairings of spin-singlet type. In fact, this model also corresponds to the mean-field theory of a generalized $t$-$V$ model, where $V<0$ represents attractive nearest-neighbor density-density interaction~\cite{RevModPhys.62.113}. Although it was suggested that the ground state of this model in the large-$V$ regime is a superconductor with a mixture of $s$- and $d$-wave pairings, the exact nature of this mixed state remains unclear.

\begin{figure}
\includegraphics[width=85mm]{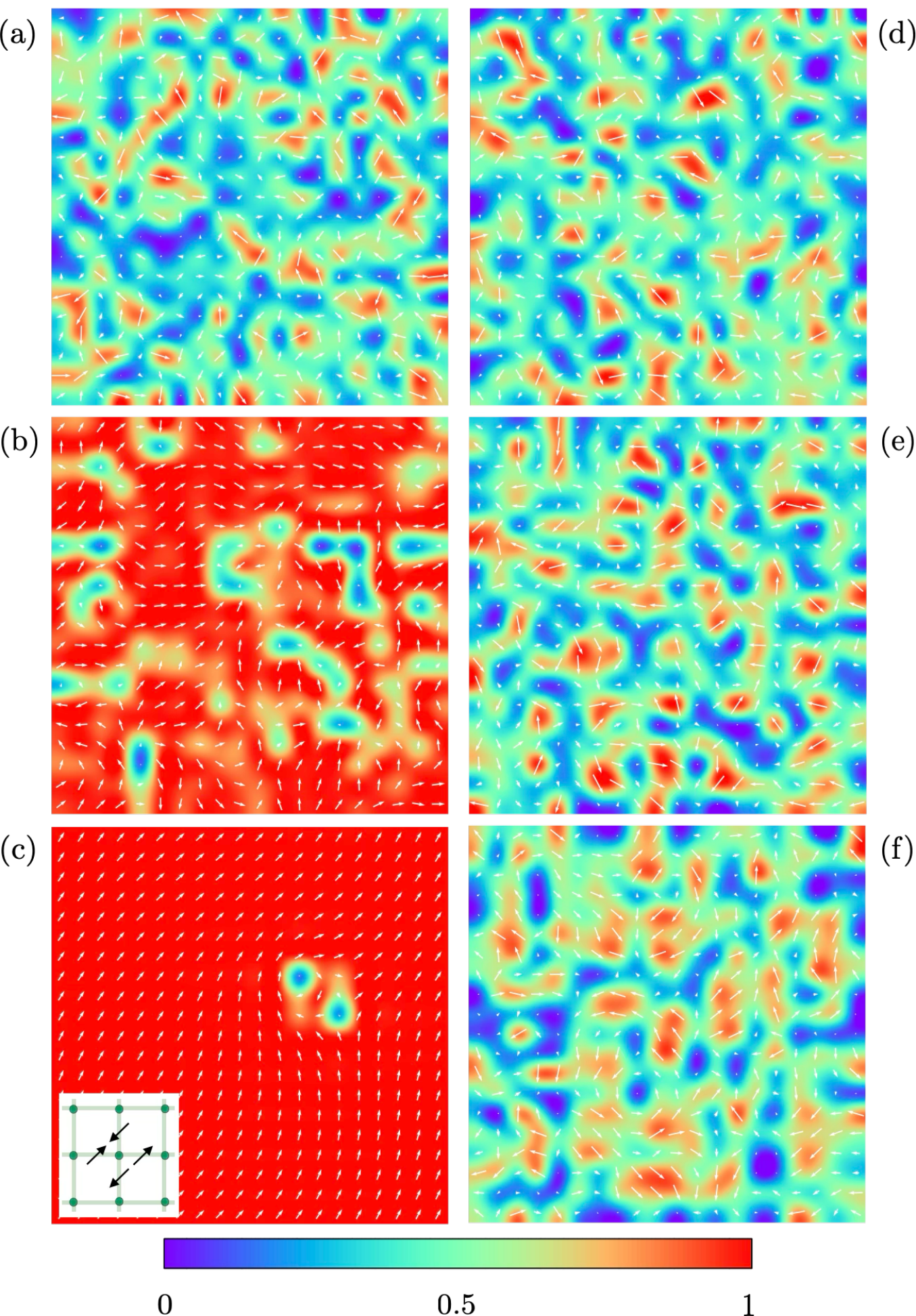}
\caption{\label{snapshots} (Color online) Snapshots of dynamical evolutions of the pairing fields, $\Delta_d$ for $g/t=0.5$ in (a) $t=0$, (b) $t=50$, and (c) $t=500$, and for $g/t=3.2$ in (d)--(f). The insert of (c) is the $d$-wave configuration. The white arrows indicate the the phase of the local $\Delta_d$ while the color density represents the corresponding amplitude $|\Delta|$.  The TDGL simulations were performed on a $40\times40$ square lattice.}
\end{figure}

We consider a model of spin-singlet pairing between nearest-neighbor electrons on a square lattice,
\begin{equation}
\setlength\abovedisplayskip{4pt}
\setlength\belowdisplayskip{1pt}
\label{eq:H0}
\mathcal{H}=-\sum_{ij \sigma}t_{ij}c^{\dag}_{i\sigma}c_{j\sigma} + g \sum_{ij}\left[ \Delta_{ij} c^{\dag}_{i\uparrow}c^{\dag}_{j\downarrow} + \Delta^{*}_{ij} c_{j\downarrow}c_{i\uparrow} \right],
\end{equation}
where the first term in our model represents the electron hopping terms with $t_{ij}=t$ when $i$ and $j$ are nearest neighbors, while the second term describes the pairing terms, both between the nearest neighbors (NNs), denoted by $\left<ij\right>$. Here $\sigma=\uparrow, \downarrow$ refers to the spins of the electrons. The pairing amplitudes are complex variables, defined as $\Delta_{ij}\equiv |\Delta|e^{i\theta_{ij}}$, where we fix the pairing magnitude at $|\Delta|=1$, and allow the phases $\theta_{ij}$ to vary, ensuring that $\Delta_{ij}=\Delta_{ji}$. The dynamics of the phases $\theta_{ij}$ are then investigated using the time-dependent Ginzburg-Landau (TDGL) equations~\cite{Hohenberg_1977} on a square lattice as follows,
\begin{equation}
\frac{\partial\Delta_{ij}}{\partial t}=-\gamma\frac{\partial\langle \mathcal{H} \rangle}{\partial\Delta^{*}_{ij}} + \eta_{ij}(t).
\end{equation}
In this model, $\gamma$ represents the damping coefficient, and $\eta(t)$ is a random force following a Gaussian distribution with zero mean and variance given by $\langle \eta_{ij}(t)\eta_{kl}(t') \rangle = 2 \gamma k_{B}T \,\delta_{ij, kl}\delta(t-t')$~\cite{Kubo_2012}.

The paring Hamiltonian in Eq.~(\ref{eq:H0}) in the small $g/t$ regime has been studied in the context of the mean-field approximation to a generalized Hubbard model with $d$-wave pairing symmetry~\cite{zhu2016bogoliubov}.  To characterize the $d$-wave {superconducting (SC)} order, two local pairing parameters $\Delta_d(\mathbf r_i) = (\Delta_1 - \Delta_2 + \Delta_3 - \Delta_4)/4$ and $\Delta_s(\mathbf r_i) = (\Delta_1 + \Delta_2 + \Delta_3 + \Delta_4)/4$ associated with a site at $\mathbf r_i$ are introduced; here $\Delta_1$, $\Delta_2$, $\Delta_3$, and $\Delta_4$ are the pairing amplitudes on the four nearest-neighbor bonds connected to site-$i$. Snapshots of the local $\Delta_d(\mathbf r)$ obtained from the TDGL simulation of a relaxation process are shown in  FIG.~\ref{snapshots} (a)-(c) for $g/t = 0.5$. In this small $g$ case, the system initially prepared in a disordered state is found to relax to a $d$-wave SC state. This ferromagnetic order of the $\Delta_d$ parameter results from a staggered arrangement of the local nearest-neighbor pairing $\Delta_{ij}$ as shown in the inset of FIG.~\ref{snapshots}(c). The XY nature of the local $d$-wave order parameter implies that the phase ordering process is characterized by the formation and subsequent pair-annihilation of vortices, which is confirmed in our simulations. 

\begin{figure}
\includegraphics[width=85mm]{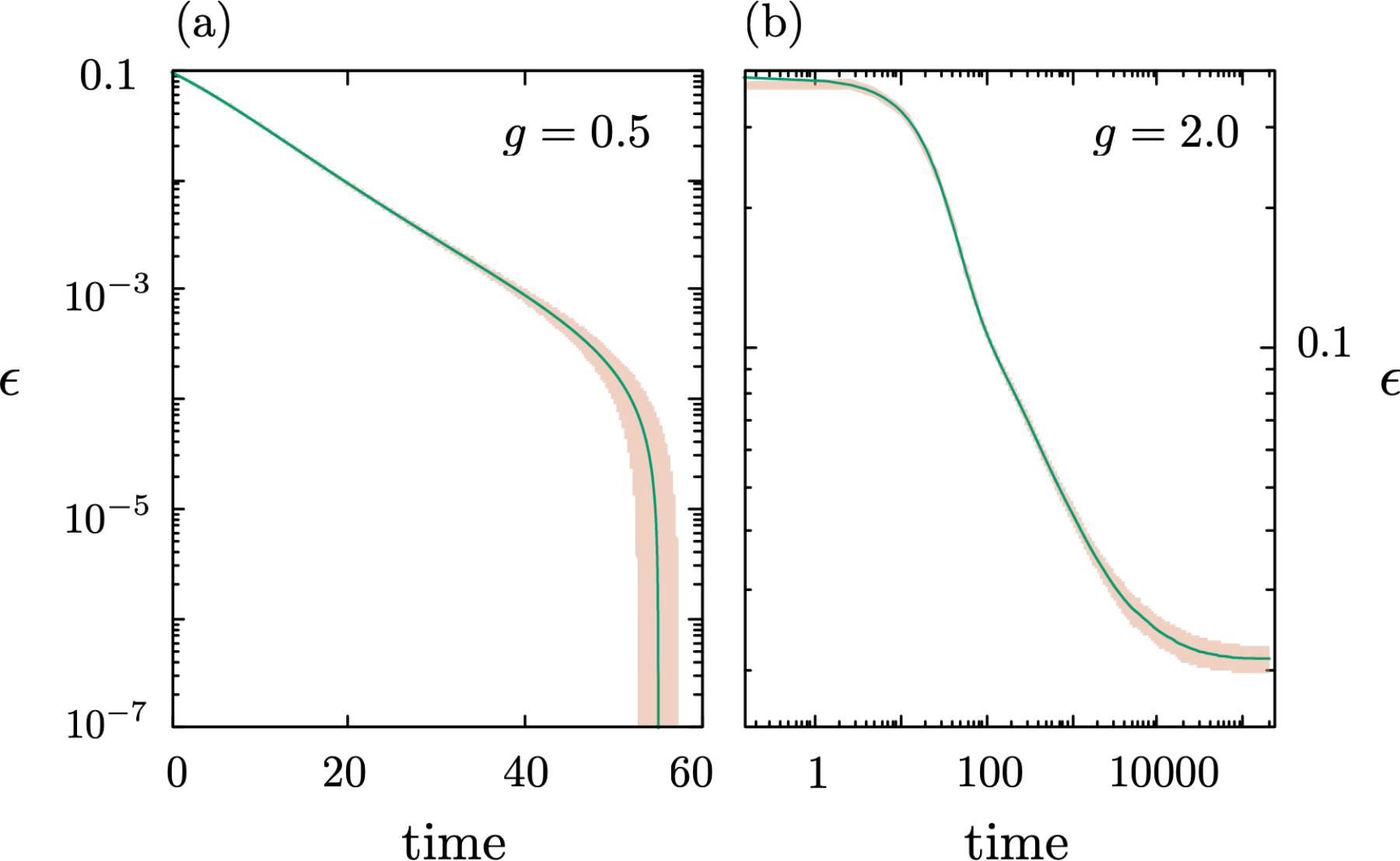}
\caption{\label{energy_time} (a) semi-log plot of the system energy density relative to the uniform $d$-wave superconducting order versus time for $g/t=0.5$, (b) log-log plot of the relative system energy density versus time for $g/t=2.0$} 
\end{figure}

However, for the case of large pairing, $g/t = 3.2$, our simulations showed that the system cannot relax to the $d$-wave SC state even after a long relaxation time. Instead, the system is trapped in a state with disordered pairing configuration as shown in FIG.~\ref{snapshots} (d)-(f). The freezing behavior in the strong pairing regime is further illustrated in FIG.~\ref{energy_time} which shows the system energy density relative to the ground state, $\epsilon(t) \equiv [E(t) - E_0] / N$, as a function of time for both small and large $g$. Here $E_0$ is the energy of the ground state which is the uniform $d$-wave SC order in the case of small $g$ phase. The determination of $E_0$ in the large-$g$ phase will be detailed below. Notably, rather distinct behaviors can be seen for the relaxation dynamics of the two pairing regimes. For $g/t=0.5$, the relaxation is characterized by an exponentially decaying energy curve $\epsilon \sim \exp(-t / \tau)$, where $\tau$ is the energy relaxation time constant, as indicated by the straight line in the semi-log plot of FIG.\ref{energy_time}(a). The fast drop of $\epsilon$ toward zero at $t \sim 60$ can be attributed to finite size effects. On the other hand, as shown in FIG.~\ref{energy_time}(b), the energy density in the case of $g/t=2.0$ remains finite for a prolonged period of time, indicating that the system is trapped in a state corresponding to a local energy minimum.

\begin{figure}
\includegraphics[width=85mm]{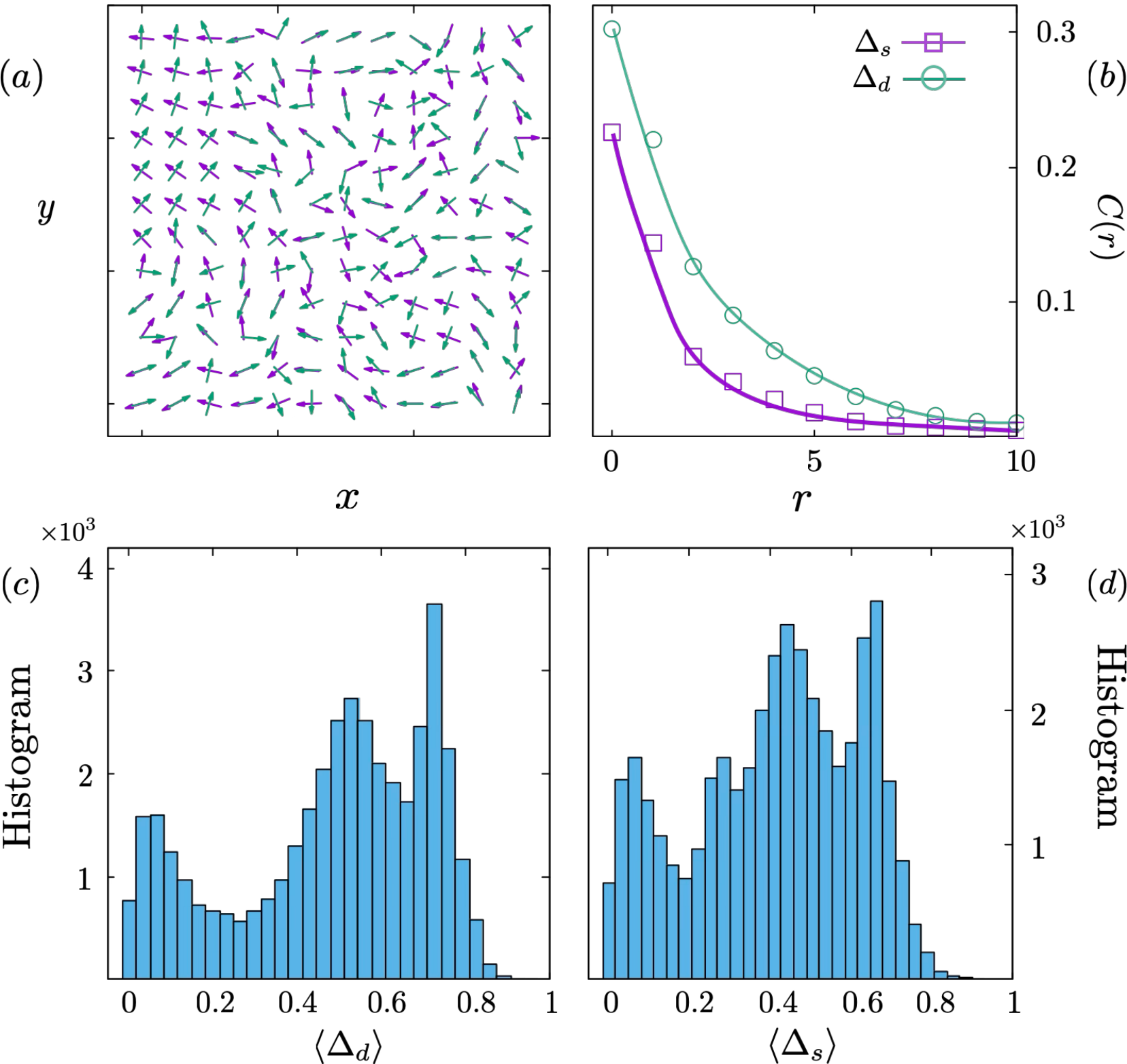}
\caption{\label{characterization} (a) A snapshot of $\Delta_s$ and $\Delta_d$ in the frozen state at late times of a TDGL simulation for $g/t=3.2$, (b) correlation function of $\Delta_s$ and $\Delta_d$, panels (c) and (d) show the histograms of amplitudes $|\Delta_d|$ and $|\Delta_s|$, respectively, in the frozen states. }
\end{figure}

Moreover, the real-space simulation results indicate these local minima are characterized by random pairing configurations. For example, FIG.~\ref{characterization}(a) shows a snapshot of the two local order parameters $\Delta_d$ and $\Delta_s$ in a frozen state at late times $\sim 10^4$ of the TDGL simulations for $g/t = 3.2$. The lack of long-range order also manifests itself in the rapid decay of correlation functions $C_{\Gamma}(\mathbf r)=\langle \Delta^{\,}_{\Gamma}(\mathbf r_0) \Delta^{*}_{\Gamma}(\mathbf r_0 + \mathbf r) \rangle$ for both pairing symmetries $\Gamma = s, d$, as shown in FIG.~\ref{characterization}(b). Moreover, not only are the phases of the local $d$ and $s$ pairing disordered, their amplitudes also exhibit strong fluctuations; see FIG.~\ref{characterization}(c)-(d) for the histogram of the amplitudes $|\Delta_d|$ and $|\Delta_s|$ obtained the frozen states of TDGL simulations.

\begin{figure}
\includegraphics[width=85mm]{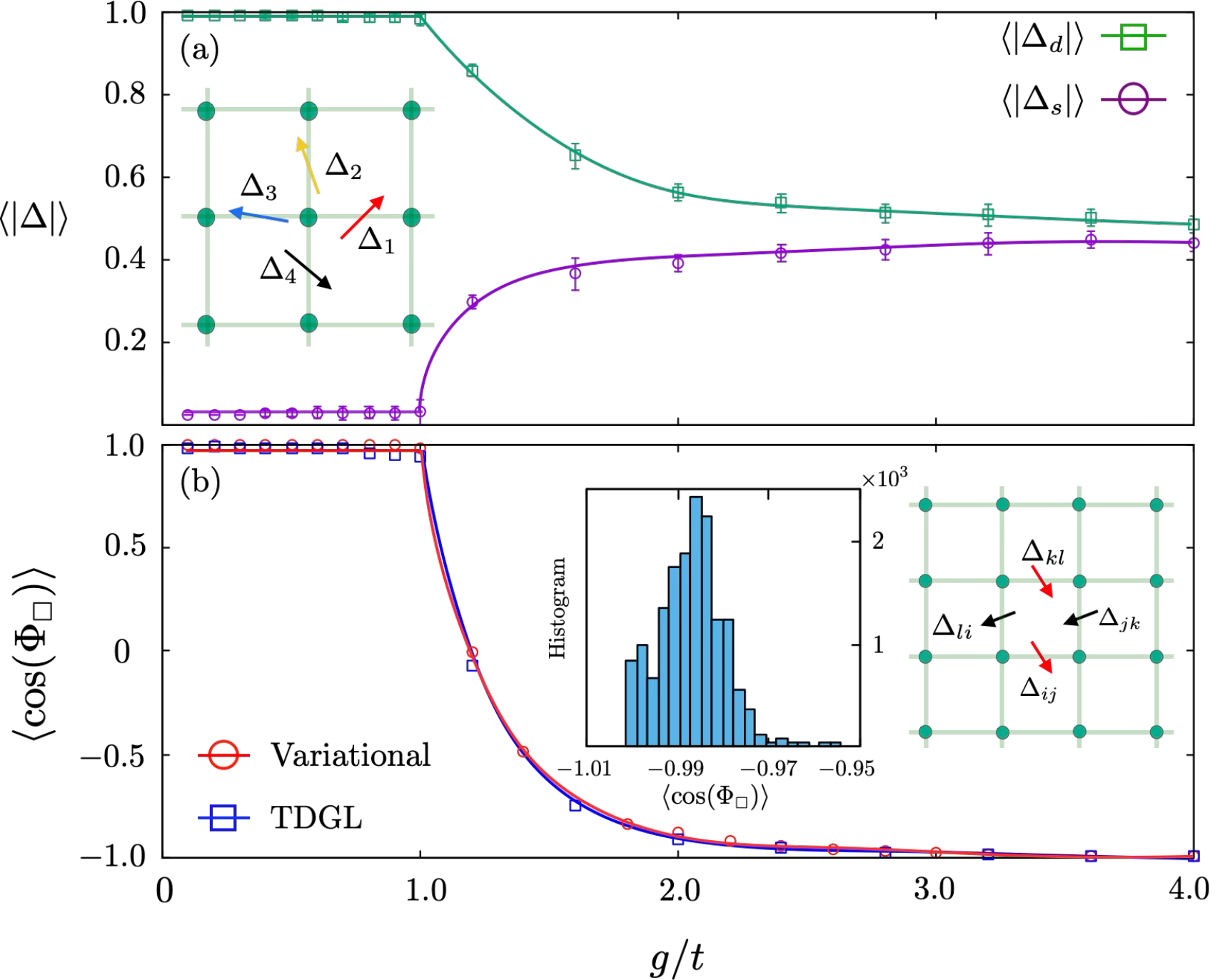}
\caption{\label{phase_diagram} (a) The amplitude of local pairing order parameters $\langle |\Delta_{s, d} | \rangle$ versus pairing strength $g/t$ in either the relaxed states (small $g$) or the frozen states (large $g$) from TDGL simulations. The insert indicates the four pairings used to define $\Delta_d$ and $\Delta_s$. (b) Averaged local flux variable $\langle \cos(\Phi_\square) \rangle$  versus $g/t$, from TDGL simulations and variational methods. The left inset shows the histogram of local fluxes in the frozen states of the large-$g$ regime, while the right inset depicts the XY spins in the orthogonal $\pi$-flux order.}
\end{figure}

To further characterize the two distinct SC behaviors discussed above, we consider the averaged amplitudes of the local SC order $\langle \left| \Delta_{d, s} \right| \rangle$, where $\langle \cdots \rangle$ indicates averaging over lattice sites and simulations with different initial conditions. These two global order parameters of either the relaxed (small-$g$) or frozen (large-$g$) states from the TDGL simulations are plotted in FIG.~\ref{phase_diagram}(a) as functions of the pairing strength. The results clearly show a phase transition at a critical pairing $g_c = t$. For small $g < g_c$, the system quickly relaxes to a uniform $d$-wave SC state characterized by $\langle {\Delta}_d \rangle = 1$ and $\langle {\Delta}_s \rangle = 0$. Above the critical point $g > g_c$, the system is found to trapped in a disordered state with both $\langle {\Delta}_{d, s} \rangle$ being nonzero. The $d$-wave order $\langle {\Delta}_d \rangle$ decreases monotonically with increasing pairing strength, while the $s$-wave pairing is gradually enhanced from zero in this strong pairing phase. It is worth noting that this mixed SC phase is consistent with previous studies showing a ``$s + d$'' SC state at the large negative $V$ regime of the generalized Hubbard model~\cite{RevModPhys.62.113}.

Although the large-$g$ phase is marked by the absence of long-range order, it nonetheless exhibits strong short-range correlations. To demonstrate this, we consider the  flux variable $\Phi_{\scriptscriptstyle\square}=(\theta_{ij}-\theta_{jk}+\theta_{kl}-\theta_{li})$ of a local square plaquette, where $i, j, k, l$ are the lattice sites at the four vertices of the plaquette, and $\theta_{ij}$ denotes the phase angle of the pairing $\Delta_{ij}$ along the edge $(ij)$, as shown in left inner panel in FIG.~\ref{phase_diagram}(b). This local variable also corresponds to the U(1) flux over a plaquette, which is invariant under gauge transformations to be discussed below. The spatially averaged plaquette flux $\langle \cos(\Phi_{\scriptscriptstyle\square}) \rangle$ in the frozen states of the TDGL simulations is shown in FIG.~\ref{phase_diagram}(b) as a function of pairing strength (the blue line). While the flux is pinned at $\langle \cos(\Phi_{\scriptscriptstyle\square}) \rangle = 1$ in the $d$-wave ground state of $g < g_c$, it decreases monotonically in the strong pairing regime and approaches $\langle \cos(\Phi_{\scriptscriptstyle\square}) \rangle = -1$ asymptotically as $g \to \infty$. 
In the large $g$ limit, the average $\pi$-flux condition actually means that the flux of individual plaquette is subject to the constraint $\cos( \Phi_{\square} ) = -1$. This is illustrated in the histogram plot of FIG.~\ref{phase_diagram}(b) which shows a very sharp peak at $\cos( \Phi_{\square} ) = -1$ for the case of $g/t = 10$. The local $\pi$-flux constraint on the otherwise disordered XY spins is reminiscent of the ice-rule type constraints in geometrically frustrated magnets.

The simplest state satisfying the local $\pi$-flux constraint is a $\mathbf q =  0$ long-range ordered state dubbed {\em orthogonal}  order.  In this ordered SC state, the XY spins on opposite edges of a square plaquette are parallel to each other, while XY spins on adjacent edges are orthogonal to each other; see the inset of FIG.~\ref{phase_diagram}(b). At large, but finite, ratio of $g/t$, we find that this orthogonal state evolves into a $\mathbf q = (\pi, \pi)$ state with a doubled unit cell. Spins in each plaquette are still close to the orthogonal configuration, but with slight deviations. Numerically, all the frozen states obtained from TDGL simulations are found to have an energy slightly greater than that of this $(\pi, \pi)$ ordered nearly orthogonal state, which is thus likely to be the ground state of the large-$g$ regime. FIG.~\ref{phase_diagram}(b) shows the cosine of the plaquette flux $\cos(\Phi_{\scriptscriptstyle\square})$ of the variationally optimized $(\pi, \pi)$ state as a function of $g/t$. We note that the fluxes obtained from ensemble-averaged TDGL simulations closely follow those of the variational $(\pi, \pi)$ state.

To shed light on the nature of the glassy states at large~$g$, we first note that the system exhibits an extensive continuous degeneracy in the $g/t \to \infty$ limit. This is because, in the absence of the hopping term, the Hamiltonian is invariant under the gauge transformation:
\begin{eqnarray}
	\label{eq:gauge-sym}
 	c^{\,}_{i,\sigma}  \to \tilde c^{\,}_{i,\sigma} = c^{\,}_{i,\sigma} \,e^{i \phi_i}, \quad
	\Delta_{ij}  \to \tilde \Delta_{ij} = \Delta_{ij}\, e^{i (\phi_i + \phi_j)}.  \quad
\end{eqnarray}
The pairing phases, or angles of XY spins, are modified as $\theta_{ij} \to \tilde{\theta}_{ij} = \theta_{ij} + \phi_i + \phi_j$. 
As a result, starting from a long-range ordered orthogonal state with $\theta_{ij} = \alpha$ for all horizontal bonds $\langle ij \rangle \parallel \hat{\bm x}$ and $\theta_{ij} = \alpha + \pi$ for all vertical bonds $\langle ij \rangle \parallel \hat{\bm y}$, one can perform a series of local gauge transforms to randomize the XY spins. Yet, the energy of the resultant disordered SC states remains the same as that of the orthogonal state because of the above gauge symmetry. Moreover, the flux variable $\cos(\Phi_{\scriptscriptstyle\square})$ can be shown to correspond to a non-Abelian flux over a Wilson loop which is gauge-invariant. This then means that  the local constraint $\cos( \Phi_{\square} ) = -1$ is exactly satisfied by all disordered states generated by gauge transformations.

\begin{figure}
\includegraphics[width=86mm]{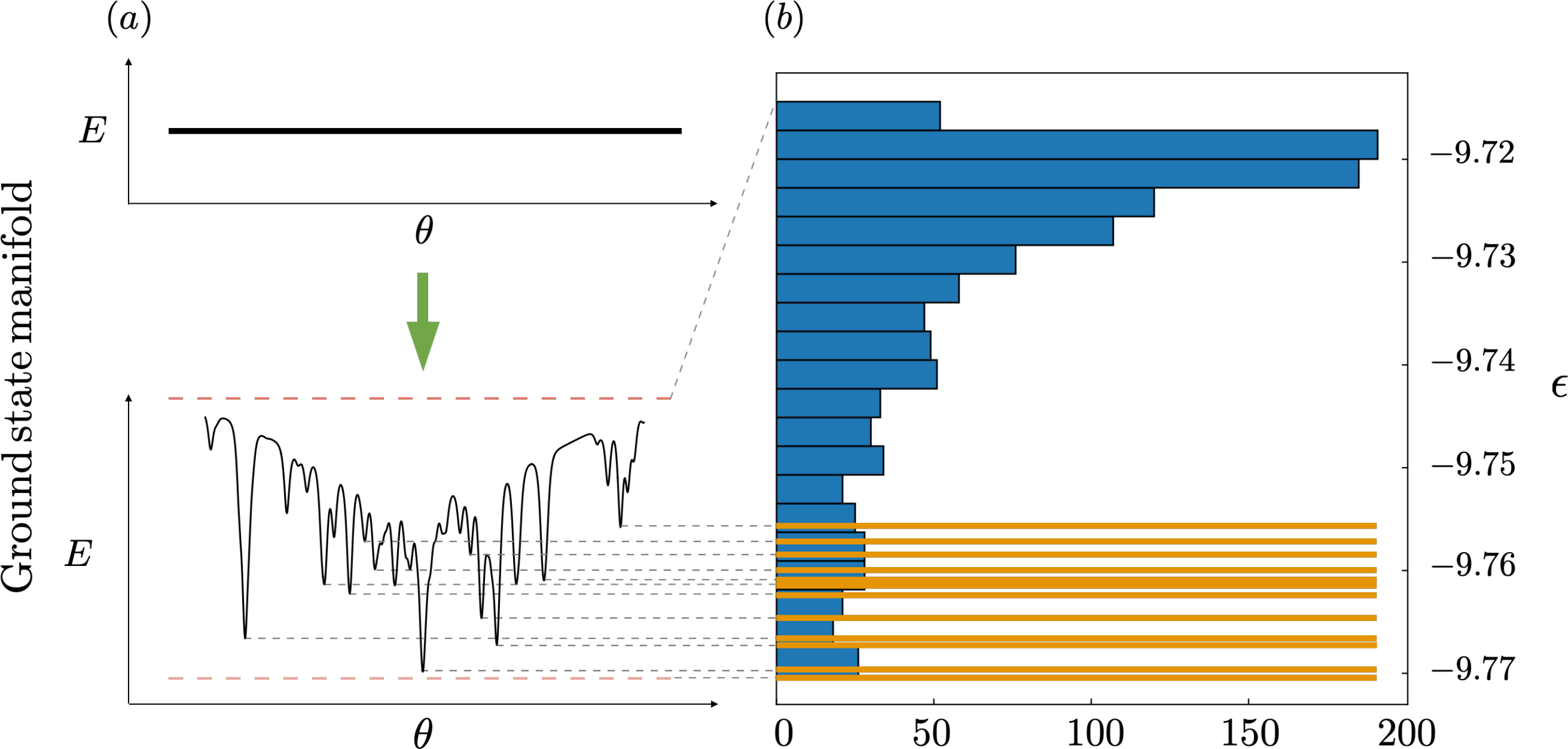}
\caption{\label{landscape} (a) A schematic diagram showing the emergence of rugged energy landscape of a glassy phase from a flat energy landscape due to continuous macroscopic degeneracy. (b) Histogram (blue boxes) of energy densities of random $\pi$-flux states, generated by local gauge transformations, in the presence of a finite hopping $t / g = 0.2$. The yellow lines indicate the frozen states obtained from TDGL simulations with an initial random $\pi$-flux state. The local $\pi$-flux constraint is still nearly satisfied in these frozen states, corresponding to local energy minima of the spontaneously-generated rugged energy landscape. }
\end{figure}

The extensive and continuous ground-state degeneracy of the system at $g \to \infty$ corresponds to a flat energy landscape in a high-dimensional configuration space. This is similar to the flat energy landscape of frustrated XY or Heisenberg spins on, e.g. kagome or pyrochlore lattices. The presence of a finite hopping spoils the gauge symmetry in Eq.~(\ref{eq:gauge-sym}), thus lifting the macroscopic degeneracy of the infinite-$g$ ground states. A scenario for the emergence of the glassy behaviors, as illustrated in Fig.~\ref{landscape}(a), is the transformation of the original degenerate manifold into a rugged energy landscape when the degeneracy is lifted. To demonstrate this scenario, we first generate $10^4$ different $\pi$-flux states through random gauge-transformations from the orthogonal order; all these states are energetically degenerate at $t = 0$. Next we turn on the hopping and compute the energies of these $\pi$-flux states using exact diagonalization; the resultant energy density histogram is shown in FIG.~\ref{landscape}(b). The finite hopping introduces a bandwidth of roughly $\Delta \epsilon \sim 0.05t$. 

Naturally most of these $\pi$-flux configurations are now excited states of the system, while the ground state is the $\mathbf q = (\pi, \pi)$ nearly orthogonal state evolved from the original orthogonal state.  Interestingly, we also find that nearly all of these $\pi$-flux states, which are related to the orthogonal states through gauge symmetry, cannot relax to the ground state. This is demonstrated by the yellow lines in FIG.~\ref{landscape}(b) which denote the energy densities of the frozen states obtained from TDGL simulations starting from randomly selected $\pi$-flux states. Importantly, we found that spins of the original $\pi$-flux states are only slightly modified during the TDGL simulations. As a result, each plaquette still maintain a flux value very close to $\pi$; also see the histogram in FIG.~\ref{phase_diagram}(b). These frozen states with disordered XY spins can thus be viewed as local energy minima emerging from the original degenerate manifold of the $g\to \infty$ limit.

To summarize, we have studied a simple disorder-free XY spin model which nonetheless exhibits a glassy phase at low temperatures in the strong-coupling regime. The scenario of a spontaneously generated spin-glass phase in the {\em absence} of quenched disorder is expected to shed a new light on the mechanisms of structural glasses in supercooled liquid, which remains an important topic in modern condensed matter physics despite many years of intensive studies. Indeed, almost all previous works on glassy states of disorder-free spin-glass models highlight the conventional view of a self-generated randomness, either in local fields or effective interactions, under the mean-field approach, which is made possible thanks to rather contrived and complicated spin interactions~\cite{bouchaud94,Marinari94a,Marinari94b,Lipowski00,Franz01,yoshino18}. On the contrary, here we present a different mechanism for the emergence of glassy states in a disorder-free system, namely the lifting of a macroscopic continuous degeneracy. Although similar pictures have previously been proposed in frustrated magnets~\cite{cepas14,klich14,yang15,mitsumoto20}, our work provides the first numerical demonstration of this mechanism in an exactly solvable model.

The itinerant-electron XY model studied here also corresponds to the mean-field theory of the SC phases of a generalized Hubbard, or $t$-$U$-$V$, model~\cite{RevModPhys.62.113}. This is one of the simplest lattice models that exhibit a SC ground state of the $d$-wave symmetry. Interestingly, the uniform $d$-wave SC order is shown to be unstable towards an intriguing SC state with both $s$ and $d$-pairing in the presence of a large density-density attraction $V$. Our work clarifies the nature of this mixed-phase state to be a special case of an emergent glassy phase.

\bigskip

\begin{acknowledgments}
{\em Acknowledgment}. This work is supported by the US Department of Energy Basic Energy Sciences under Contract No. DE-SC0020330. L. Yang acknowledges the support of Jefferson Fellowship by the Jefferson Scholars Foundation. The authors also acknowledge the support of Research Computing at the University of Virginia.
\end{acknowledgments}

\bibliography{ref}
\end{document}